\documentclass[a4paper,UKenglish]{lipics-v2021} %This is a template for producing LIPIcs articles. 
\usepackage{amsmath, amssymb}
\usepackage{microtype}
\usepackage[dvipsnames]{xcolor}
\usepackage{caption}

\usepackage{tikz}

 \newcommand{\defn}[1]{\textbf{\textit{#1}}}
\newcommand{\rank}[1]{\text{\textsc{Rank}$(#1)$}}
\newcommand{\select}[1]{\text{\textsc{Select}$(#1)$}}

\renewcommand{\paragraph}[1]{\smallskip\textbf{#1}}

\pdfoutput=1 %uncomment to ensure pdflatex processing (mandatatory e.g. to submit to arXiv)
\hideLIPIcs  %uncomment to remove references to LIPIcs series (logo, DOI, ...), e.g. when preparing a pre-final version to be uploaded to arXiv or another public repository

\title{SPIDER: Improved Succinct Rank and Select Performance} 

 \author{Matthew D. Laws}{Williams College Computer Science, Williamstown, MA USA}{mdl4@williams.edu}{https://orcid.org/0009-0000-1653-9332}{}%
\author{Jocelyn Bliven}{Williams College Computer Science, Williamstown, MA USA}{jmb13@williams.edu}{https://orcid.org/0009-0009-6856-0972}{}%
 \author{Kit Conklin}{Williams College Computer Science, Williamstown, MA USA}{kec2@williams.edu}{https://orcid.org/0009-0001-9449-0025}{}%
 \author{Elyes Laalai}{Williams College Computer Science, Williamstown, MA USA}{el12@williams.edu}{https://orcid.org/0009-0002-5311-1970}{}%
\author{Samuel McCauley}{Williams College Computer Science, Williamstown, MA USA}{srm2@williams.edu}{https://orcid.org/0000-0001-8196-9662}{}%
\author{Zach S. Sturdevant}{Williams College Computer Science, Williamstown, MA USA}{zss1@williams.edu}{https://orcid.org/0009-0009-4889-2750}{}%

\authorrunning{M. D. Laws, J. Bliven, K. Conklin, E. Laalai, S. McCauley, Z. S. Sturdevant} 

\EventEditors{Leo Liberti}
\EventNoEds{1}
\EventLongTitle{22nd International Symposium on Experimental Algorithms (SEA 2024)}
\EventShortTitle{SEA 2024}
\EventAcronym{SEA}
\EventYear{2024}
\EventDate{July 23-26, 2024}
\EventLocation{Vienna, Austria}
\EventLogo{}
\SeriesVolume{301}
\ArticleNo{22}

\Copyright{Matthew D. Laws, Jocelyn Bliven, Kit Conklin, Elyes Laalai, Samuel McCauley, Zach S. Sturdevant} 


 %TODO mandatory; please add comma-separated list of keywords

\category{} 

\relatedversion{} %optional, e.g. full version hosted on arXiv, HAL, or other respository/website
\supplement{}
\supplementdetails[subcategory={}, cite={}, swhid={}]{Software}{https://github.com/williams-cs/spider}

\funding{This research was supported in part by NSF grant CCF 2103813.}%

\nolinenumbers %uncomment to disable line numbering

\begin{document}

\maketitle

 \begin{abstract} 
 Rank and select data structures seek to preprocess a bit vector to quickly answer two kinds of queries: $\rank{i}$ gives the number of $1$ bits in slots $0$ through $i$, and $\select{j}$ gives the first slot $s$ with $\rank{s} = j$.  A succinct data structure can answer these queries while using space much smaller than the size of the original bit vector.

 State of the art succinct rank and select data structures use as little as 4\% extra space (over the underlying bit vector) while answering rank and select queries very quickly. Rank queries can be answered using only a handful of array accesses.  Select queries can be answered by starting with similar array accesses, followed by a linear scan through the bit vector.

 Nonetheless, a tradeoff remains: data structures that use under 4\% space are significantly slower at answering rank and select queries than less-space-efficient data structures (using, say, over 20\% extra space).

 In this paper we make significantly progress towards closing this gap.  We give a new data structure, SPIDER, which uses 3.82\% extra space.  SPIDER gives the best known rank query time for data sets of 8 billion or more bits, even compared to much less space-efficient data structures.  For select queries, SPIDER outperforms all data structures that use less than 4\% space, and significantly closes the gap in select performance between data structures with less than 4\% space, and those that use more (over 20\% for both rank and select) space.

 SPIDER makes two main technical contributions.  For rank queries, it improves performance by interleaving the metadata with the bit vector to improve cache efficiency.  For select queries, it uses predictions to almost eliminate the cost of the linear scan.  These predictions are inspired by recent results on data structures with machine-learned predictions, adapted to the succinct data structure setting.  Our results hold on both real and synthetic data, showing that these predictions are effective in practice.
\end{abstract}

%SEA allows title page
\newpage
\setcounter{page}{1}

\section{Introduction}%
\label{sec:intro}

 Rank and select are classic bit vector queries in computer science.  Given a bit vector $V$, $\rank{i}$ gives the number of $1$ bits among $V[0],\ldots, V[i]$.  $\select{j}$ gives the location of the $j$th $1$ bit; in other words, $\select{j}$ gives the smallest $i$ such that $\rank{i} = j$.  
Thus, if $n$ is the length of $V$, and $n_1$ is the number of $1$ bits it contains,
rank queries are well-defined for $i = 0, \ldots, n-1$ and select queries are well-defined for $i = 1,\ldots, n_1$.  

Rank and select data structures are most interesting when they are restricted to small space.  After all, with unlimited space one could just store the answer to all $n + n_1$ rank and select queries.  

Thus, we focus on \defn{succinct} data structures: those that use much less than $n$ bits of space in addition to $V$.

Rank and select queries can be used to implement many other succinct data structures, 
including, for example,
trees and graphs~\cite{MunroRaman97,RamanRaSa07}, 
suffix trees~\cite{MakinenNavarro05}, 
and filters~\cite{PandyBeJo17b,LiuYiYu20}.
 Wavelet trees are a particularly motivating example, as they use a sequence of rank and select data structures as a subroutine~\cite{Navarro14,GrossiGuVi03}, and have applications in areas such as text compression, DNA alignment, and computational geometry~\cite{DinklageElFi21}.
We note that all of these applications require a data structure that can handle \emph{both} rank and select queries on $V$.

These applications have motivated a long line of work on engineering rank and select data structures that can answer queries quickly with very small space overhead~\cite{ZhouAnKa13,Vigna08,Jacobson89,Munro96,GonzalezGrMa05,NavarroProvidel12}.  When space is at a premium, the state of the art result is \texttt{pasta-flat} by Kurpicz~\cite{Kurpicz22}.  His data structure uses only 3.58\% extra space ($0.0358n$ bits in addition to $|V|$), and is very performant, achieving query time up to 16.5\% faster than the previous state of the art.

However, rank and select query performance still incurs a tradeoff between query time and space.  For example, Vigna~\cite{Vigna08} gives a data structure for rank queries using 25\% space that is roughly 19\% faster than \texttt{pasta-flat}, and a data structure
 for select queries using 12.2\% space, which is roughly 65\% faster than \texttt{pasta-flat}.  

In this paper, we make significant progress on overcoming this tradeoff.

 \subsection{How Rank and Select Data Structures Work}%
\label{sec:highlevelsummary}

We give a high-level description of the basic ideas behind rank and select data structures.  This is a useful introduction to our data structure, and gives context to related work.  We begin with the naive data structure: one can store the solution to all rank and select queries, requiring $64(n + n_1)$ bits of space.  Throughout the paper we give data structures that allow for $n$ up to $2^{64}$, and thus each bit vector index requires $64$ bits.

\subsubsection{Searching in a Cache Line}
\label{sec:lastmile}

 Rank and select data structures often operate on bit vectors with billions of bits.  Therefore, cache efficiency is by far the most important consideration for performance.  
 In this paper we assume a cache line size of 512 bits and a word size of 64 bits.  
 (Our results likely generalize by adjusting the parameters in Section~\ref{sec:datastructure}.)

 With cache efficiency in mind, we can improve the naive rank and select data structure.  To begin, we observe that there is no need to store metadata to help with search within a cache line.  This is because cache efficiency is our primary goal: a single cache miss allows us to bring in $512$ bits, and look for $1$s manually.  First we describe how this can be used to save space, and second we describe how to decrease the computation cost within a cache line.  This is the \defn{strawman} data structure, also described in~\cite{ZhouAnKa13}.

\paragraph{Saving Space.}  The strawman data structure has a \defn{rank array} of length $\lfloor n/512\rfloor$ that stores the number of $1$s before each cache line: its $i$th entry stores $\rank{512i - 1}$.  To answer $\rank{i}$, first look up the $\lfloor i/512\rfloor$th entry in the rank array and store its value in $r_1$ (thus, $r_1$ is the number of $1$ bits in slots $0$ through $512\lfloor i/512\rfloor - 1$ of $V$).  Then, access $V$ to find the number of $1$s in slots $512\lfloor i/512\rfloor$ through $i$; adding this to $r_1$ gives $\rank{i}$.

For select, 
the strawman data structure
uses a heuristic 
common to most succinct data structures,
called \defn{rank-based select}~\cite{Kurpicz22}.  
First, set a parameter $\sigma$ (often $\sigma = 8192$).
Similar to rank, the strawman data structure stores the \defn{select array}, whose $i$th slot stores $\select{\sigma\cdot i + 1}$, for $i = 0,\ldots \lfloor(n_1-1)/\sigma\rfloor$.

The idea of rank-based select is to use the select array to get an approximate solution, and then refine it with a linear search---but we note that using the rank array saves considerable time for this search.  To find $\select{i}$, use the select array to look up $s = \select{\lfloor (i-1)/\sigma\rfloor+1}$.  Then, begin a linear scan in the rank array from position $\lfloor s/512\rfloor$; the goal is to find the largest entry in the rank array that is smaller than $i$.  Let $e$ be this entry and $p$ be its position in the rank array; once $e$ and $p$ are found, search for the $(i-e)$th $1$ bit within the cache line starting at $512p$ in $V$.

We note that rank-based select is $\Omega(n)$ time in the worst-case due to the linear scan, but past work has shown that it is efficient in practice~\cite{Kurpicz22,ZhouAnKa13}.

The strawman data structure uses, for $\sigma = 8192$, $64\cdot n/512 + 64\cdot n_1/8192 \leq .133n$ space (13.3\% overhead), and is very cache-efficient.  However, the cost for a naive scan within a cache line is considerable.  Fortunately, as we now explain, many modern machines support low-level operations that make this cost almost negligible.

\paragraph{Rank Search Within a Cache Line.}  Many modern processors have a \texttt{popcount} operation that gives the number of $1$s in a 64-bit word.  Thus, we can find the number of $1$s in a cache line at or before position $i$ with at most $8$ \texttt{popcount} operations.  Before the final popcount operation, we must right shift to remove all bits in the word after $i$.  This results in an extremely computationally efficient way to compute the rank within a cache line.

\paragraph{Select Search Within a Cache Line.}  Pandey, Bender, and Johnson~\cite{PandeyBeJo17a} give a method to search within a bit vector using a clever application of specialized processor instructions.  In particular, many modern x86 machines support the \texttt{pdep} and \texttt{tzcnt} instructions.  A single application of these two instructions allows us to find the location of the $i$th $1$ bit in a $64$-bit word; see~\cite{PandeyBeJo17a} for details.
 Thus, we can find the $j$th $1$ bit within a cache line as follows.  We perform up to $8$ \texttt{popcount} instructions, keeping a running count $c$ of the number of $1$s seen.  If adding the \texttt{popcount} of the new word would cause $c > j$, we instead use \texttt{pdep} and \texttt{tzcnt} to find the $(j-c)$th $1$ bit within the word.

This methodology is often called \defn{fast select}, and it noticeably speeds up select performance on the machine we used in our experiments (see, for example, the difference between \texttt{pasta-flat} and \texttt{pasta-flat-fs} in Figure~\ref{fig:select}).  
We note that, except for \texttt{pasta-flat} which we keep as a baseline, all data structures we compare to in our select experiments use fast select within a cache line.

\subsubsection{Multi-Level Rank Data Structures}
To reduce space past 13\%, many data structures use a \defn{multi-level rank data structure}.

Consider splitting the bit vector into blocks of size $2^b$ (we will retain a linear search in the cache line, so assume $2^b \gg 512$).  Then, we can store two rank arrays. The $i$th entry in the high-level rank array stores the number of ones before bit $2^b\cdot i$ in $V$.  The low-level rank array recurses within a $2^b$-size block: the $j$th entry of the low-level rank array stores the number of ones between bit $2^b\cdot \lfloor 512\cdot j/2^b\rfloor$ and bit $512j$ of $V$.  

To find $\rank{i}$ we sum the $\lfloor i/2^b\rfloor$th entry in the high-level array, the $\lfloor i/512\rfloor$th entry in the low-level array, and finally a scan within the cache line beginning at slot $512\lfloor i/512\rfloor$ of $V$.  Select searches can be handled similarly: first a linear scan is performed in the high-level rank array, and then in the low-level rank array, finally searching within a cache line.

The multi-level approach incurs an extra array lookup in exchange for less space.  The top-level array requires space $64\cdot n/2^b$.  Each entry in the second-level array is $< 2^b$ and can therefore be stored in $b$ bits, giving space $b\cdot n/512$.  
 The most performant and space-efficient data structures recurse another time, using three rank arrays~\cite{ZhouAnKa13,Kurpicz22}.  
 
\paragraph{Improving Select Queries with Multi-Level Rank.}
The multi-level rank data structure improves select queries as well.  For example, consider a select query where $\select{\sigma\lfloor i/\sigma\rfloor}$ is $10\cdot 2^b$ slots away from $\select{i}$.  With a single rank array, the algorithm would have to scan through roughly $10\cdot 2^b/512$ array slots.  Meanwhile, with the two-level data structure, the algorithm would scan through roughly $10$ entries in the top-level array, followed by $2^b/512$ entries in the low-level array; for large $b$ nearly a factor $10$ speedup.  

This speedup leads to a tension between rank and select queries.  Optimizations to the rank data structure are limited by their impact on select performance.  This tradeoff can be seen in the \texttt{pasta-wide} data structure of Kurpicz~\cite{Kurpicz22}. \texttt{pasta-wide} is a simple two-level rank data structure with extremely strong performance; unfortunately, the simplified rank array makes select queries extremely slow.  This holds even if the select queries use binary search in an attempt to speed up performance.  

One of the main contributions of SPIDER is to resolve this tension: we show how \defn{predictions} can reduce the cost of a select query even for highly efficient rank data structures, achieving the best of both worlds.

\paragraph{Interleaving Rank Arrays.}  For sufficiently large bit vectors, accessing each rank array may incur a separate cache miss, significantly impacting performance.  For this reason, Zhou et al.~\cite{ZhouAnKa13} introduced the idea of \defn{interleaving} the rank arrays.  
 Specifically, in the above description, we access entry $k$ of the high-level rank array only when querying $\rank{i}$ with $k 2^b\leq i < (k+1)2^b$; thus
 the entry $j$ accessed in the low-level rank array satisfies $j\in \{k2^b/512, \ldots, (k+1)2^b/512-1\}$.  Thus, Zhou et al.\ store a single array storing (for each $k$) the $k$th entry of the high-level array, followed by all possible lower-array entries given above.  If these entries all fit in $512$ bits, both arrays can be accessed with a single cache miss.

 \subsection{Our Contributions}

We give a data structure, SPIDER (a \textbf{S}uccinct \textbf{P}redictive \textbf{I}n\textbf{DE}x for \textbf{R}ank and select) which improves rank and select performance.  

For rank queries, SPIDER uses a simple two-level data structure---but interleaves the lower-level rank array \emph{with the bit vector itself}.  Thus, if the high-level rank array fits in cache, SPIDER can answer rank queries with a single cache miss.  This noticeably improves rank query time, even compared to data structures that use far more space.  

For select queries, SPIDER avoids a costly scan using \defn{predictions}.  In particular, we draw inspiration from recent work on learned data structures, e.g.~\cite{KraskaBeCh18, FerraginaVinciguerra20}.  Learned data structures use best-fit lines to predict the answer to queries; past work has shown that real world data is roughly piecewise linear, leading to good results.  
SPIDER applies this principle to select queries.  We assume that successive select array entries are separated by approximately-evenly-spaced $1$s, and use this assumption to warm-start our linear scan.  

SPIDER also uses the multi-level strategy on the \emph{select} array to allow us to sample more bits, further improving select performance; to our knowledge this is the first data structure to use a multi-level select array.

We also give a variant, Non-Interleaved SPIDER, which does not interleave the rank array with the bit vector.  Non-Interleaved SPIDER's rank performance is worse than SPIDER but competitive with the state of the art; its select performance is similar to SPIDER.

Overall, SPIDER gives improved bounds over any known succinct data structure for {rank} queries.  For {select} queries, SPIDER improves on the state of the art data structure with less than 5\% extra space usage, and nearly matches the performance of the best known succinct data structure that uses 12.5\% space and cannot answer rank queries.  In particular, compared to the best known data structure with under 5\% space, SPIDER gives up to a 22\% speedup for rank, and 41\% speedup for select.

\subsection{State of the Art Rank and Select}
\label{sec:related_state_of_the_art}
\begin{table}[t]
    \centering
     \begin{tabular}{c|c c c c} 
         Structure & Cite & Space & Rank(ns) & Select(ns)
         \\ [0.5ex] 
         \hline
         \rule{0pt}{2.5ex}
        \texttt{vigna-rank} & \cite{Vigna08} & 25.00 & 40.69 & 179.15 \\
        \texttt{sdsl-v} & \cite{GogBeMo14} & 25.00  & 40.31 & - \\
        \texttt{pasta-wide} & \cite{Kurpicz22} & 3.23 & 38.81 & - \\
        \texttt{vigna-select} & \cite{Vigna08} & 12.20 & - & 95.04 \\
        \texttt{vigna-select-H} & \cite{Vigna08} & 15.63 & - & \textbf{71.56} \\
        \texttt{pasta-flat-fs} & \cite{Kurpicz22} & 3.58 & 50.20 & 207.47  \\
        \texttt{ni-spider} & \S\ref{sec:ni-spider} & 3.62 & 36.98 & 138.61 \\
        \texttt{spider} & \S\ref{sec:datastructure} & 3.83 & \textit{\textbf{33.56}} & \textit{126.96}  \\
     \end{tabular}
    \caption{Comparing the performance of succinct rank and select data structures.  See Section~\ref{sec:experiments} for details of the experiments.  We list the space used as a percentage of the size of the original bit vector, and the average time for rank and select on a 8 billion bit segment of the Wikipedia dataset.  The best performance numbers are in bold, and the best among data structures with under 5\% space are italicized.}
    \label{table:dataset features}
\end{table}

We use the following rank and select data structures as a baseline to compare our data structure.  
See Table~\ref{table:dataset features} for a summary of the advantages of each.

\texttt{poppy}: 
Uses a three-level strategy, interleaving the last two levels, with carefully implemented bit tricks to improve rank and select performance~\cite{ZhouAnKa13}.

The publicly-available implementation contains a bug that results in incorrect queries for bit arrays beyond 4 billion bits. 

\texttt{pasta-flat}:
Introduced by Kurpicz~\cite{Kurpicz22}, \texttt{pasta-flat} achieves the state of the art performance for rank and select data structures with less than 5\% space overhead.
 The basic idea of \texttt{pasta-flat} is a three level data structure with interleaved second and third levels, much like \texttt{poppy}.  However, \texttt{pasta-flat} uses a different strategy for interleaving the bottom two arrays using $128$-bit entries.  These entries are handled efficiently using SIMD operations.  Overall, Kurpicz showed that \texttt{pasta-flat} gave a significant speedup over \texttt{poppy} (therefore, we only compare to \texttt{pasta-flat} in our tests).

 One advantage of \texttt{pasta-flat} is that it supports both select$_1$ queries (as defined in this paper), as well as select$_0$ queries, which give the location of the $j$th $0$.  We do not consider select$_0$ queries; however, SPIDER can be easily extended to handle select$_0$ queries by using a second select data structure (that stores the location of sampled $0$s rather than sampled $1$s), slightly increasing the space.

\texttt{pasta-flat-fs}: Identical to \texttt{pasta-flat}, but uses fast select~\cite{PandeyBeJo17a} within each cache line.

\texttt{pasta-wide}:  The second \texttt{pasta} data structure~\cite{Kurpicz22} uses a simple two-level structure for very fast rank queries.  Adding the select structure from \texttt{pasta-flat} results in large space and slow select performance; therefore we only consider rank queries on \texttt{pasta-wide}.

\texttt{vigna-rank}:
Vigna's rank structure, rank9, follows a two level approach~\cite{Vigna08}.  Entries in the two levels are grouped into 128-bit ``words'' that are unpacked during select queries using broadword programming techniques.  
Rank9 has a select data structure, select9, but we do not include it in Section~\ref{sec:experiments} because it is much slower than \texttt{vigna-select}. 

\texttt{vigna-select}:
Vigna also proposes a select-only structure (called ``simple select'' in~\cite{Vigna08})  which we refer to as \texttt{vigna-select}. 
\texttt{vigna-select} is a position based select structure (it is not rank based); see~\cite{Vigna08} for more details. This data structure takes a tuning parameter; we use $2$ as it gave the best performance (see also the comparison in~\cite{Kurpicz22}).

\texttt{vigna-select-H}: uses similar techniques to \texttt{vigna-select} but is specifically optimized for bit vectors where each bit is a $1$ with probability $.5$.

\texttt{sdsl-v}:
The Succinct Data Structure Library~\cite{GogBeMo14} contains an implementation of the \texttt{vigna-rank} data structure described above.

\subsection{Other Related Work}
There are tight upper and lower theoretical bounds for rank and select data structures~\cite{Patrascu08,LiLiYu23}.

\paragraph{Compressed Bit Vectors.}
An orthogonal line of work has looked at how to save space by compressing the underlying bit vector $V$.  In particular, many practical bit vectors can be stored using much less than $n$ bits of space. The goal of this line of work is a trade-off between how much space is used on practical bit vectors, and how much time rank and select queries take.  See for example~\cite{RamanRaSa07,OttavianoVenturini14,Navarro16,BoffaFeVi22}.

There are two downsides to compressing $V$. First, this compression is situational: some bit vectors have high entropy and cannot be compressed. Second, queries require uncompressing the underlying data, which results in a cost overhead; see e.g.\ the experiments in~\cite{BoffaFeVi22}.

In this paper we assume that $V$ is stored uncompressed.  It is plausible that our ideas could be combined with past work to speed up compressed queries as well.

\paragraph{Learned Data Structures.}
 A recent, exciting line of work has looked at \defn{learned data structures}~\cite{KraskaBeCh18,Mitzenmacher18,FerraginaVinciguerra20}. Rather than keeping a worst case index, learned data structures use machine learning techniques to store high-level information about the data.  

More recently, Boffa, Ferragina, and Vinciguerra~\cite{BoffaFeVi22} used a learned index to achieve an improved data structure for rank and select. Their focus was distinct from ours: their goal was to compress the bit vector itself (as mentioned above), whereas our goal is to store succinct metadata to speed up queries.

However, SPIDER's predictive method for select uses a structure that is reminiscent of these learned indices. In particular, the basic strategy of learned indices is to a store a sequence of best-fit lines on the underlying data.  Past results indicate that on real world data, best fit lines can often store much more accurate and much more concise information than classic worst-case data structures~\cite{KraskaBeCh18,Mitzenmacher18,DingMiYu20,FerraginaVinciguerra20}.

One can view our select methodology as a very lightweight application of these learned indices. We do not have space to explicitly store best-fit lines; instead, we look up the number of ones in a subarray to roughly estimate what a best fit line could look like for a given subsection of the bit vector.  This allows us to estimate the best place to start when scanning the bit vector during a select query.

\section{The SPIDER Data Structure}
\label{sec:datastructure}

We now describe our data structures: first SPIDER, and then Non-Interleaved SPIDER.  Then we compare the techniques of SPIDER and Non-Interleaved SPIDER and how these differences impact performance.

\begin{table}[t]
\begin{center}
\begin{tabular}{ c l }
\hline
$V$ & The bit vector we are querying. \\
$n$ & Length of the bit vector. \\ 
$n_1$ & Number of 1 bits in the bit vector. \\ 
$i$ & A given query (Rank or Select). \\ 
$s$ & The predicted superblock index.\\
$r_1$ & Number of 1 bits in all slots before the superblock containing the query. \\
$b_2$ & Index of the basic block containing the query. \\
$r_2$ & 16 bit count prepended to the original block; represents the number of 1 bits \\ & 
from a the start of the superblock until the start of the basic block. \\
$\sigma_h$ & High-level sampling threshold: the frequency with which locations are stored \\ & in the high-level select array.\\

$\sigma_\ell$ & Low-level sampling threshold: the frequency with which locations are stored \\ &  in the low-level select array.\\

$\ell$ & Index in the low-level select array; gives a lower bound on the query solution. \\
$p$ & Prediction for a select query. \\
$p'$ & Altered prediction when the sampled bits cross a superblock boundary. \\
$B$ & First basic block we search in given $p$ or $p'$.\\
$\sigma$ & Sampling threshold for Non-Interleaved SPIDER.\\
 \hline
\end{tabular}
\end{center}
    \caption{Table of notation used in this paper.}
    \label{tab:notation}
\end{table}

\subsection{SPIDER Rank}
\label{sec:spider_rank}

\paragraph{SPIDER Rank Data Structure.}

SPIDER stores two arrays to help answer rank queries: a rank array, and a modified bit vector.

 First, consider partitioning the bit vector $V$ into $n/63488$ contiguous \defn{superblocks} of $63488$ bits\footnote{We assume for simplicity that the bit vector is padded to have length a multiple of $63488$.} (we explain this constant below).  
Thus, we say that superblock $i$ \defn{contains} a slot $s$ if $63488i \leq s < 63488(i+1)$. 
 The \defn{rank array} stores the number of $1$s \emph{before} the first bit in each superblock.\footnote{The first entry in the rank array is always $0$, but we store it anyway for simplicity.  Recall that $\rank{i}$ is defined to be inclusive of $i$---thus, the number of bits before slot $i$ is $\rank{i-1}$.}
 Thus, the rank array consists of $n/63488$ $64$-bit entries and the $i$th entry in the rank array stores $\rank{63488i-1}$.  
 
 Then, we store a \defn{modified bit vector}.  
 We partition $V$ into $n/496$ contiguous \defn{original blocks} of $496$ bits.  
Thus, each superblock consists of 128 original blocks.  
 The modified bit vector consists of $n/496$ \defn{basic blocks}, each of $512$ bits.
 For original block $i$ starting at slot $s_o$, let $s_b$ be the first slot in the superblock containing $s_o$; we define the \defn{local rank} to be the number of $1$s in slots $s_b$ through $s_o - 1$ (inclusive).

 Thus, the local rank of original block $i$ is $\rank{496i - 1} - \rank{63488\lfloor 496i/63488\rfloor - 1}$.
The $i$th basic block has two parts: first, a $16$-bit number storing the local rank of the $i$th original block, followed by the $496$ bits of the $i$th original block.  The modified bit vector consists of all basic blocks.

After the modified bit vector has been constructed, we no longer need to store $V$.  (Its contents are present in the modified bit vector anyway.)  Thus, our data structure consists only of the rank array and the modified bit vector.

\paragraph{Preprocessing.}
The rank array and the modified bit vector can be created simultaneously during a linear scan over the original bit vector.

\paragraph{Queries.}
On a query \rank{i}, we begin by using the rank array to find $r_1 = \rank{63488\lfloor i/63488\rfloor - 1}$ (this can be done by looking up the $ \lfloor i/63488 \rfloor$th position in the rank array). 

Then, we access the modified bit vector. Bit $i$ can be found in the $b_2 = \lfloor i/496 \rfloor$th basic block.
Let $r_2$ be the 16-bit value stored at the beginning of block $b_2$ in the modified bit vector.  Then $r_2 + r_1$ gives the rank of the last slot before $b_2$.

Finally, we must count the number of $1$s before position $i$ within $b_2$.  
This can be found using the \texttt{popcount} instruction on the first $\lfloor i/64\rfloor + 1$ words of the $8$ words in the basic block.  During this process, we must mask out the first $16$ bits of the first word, and then remove all bits after $i$ in the last word using a shift operation.  Summing the result of all of these \texttt{popcount} operations with $r_2 + r_1$ gives the rank of $i$.

\paragraph{How we Chose the Superblock Size.}

The key parameter we chose is the size of the local rank; we chose a moderate value of $16$ bits.  If the local rank fits in $16$ bits, each superblock must have at most $2^{16}$ slots.  The basic block is the size of a cache line ($512$ bits), and $16$ bits are used to store the local rank, leaving $496$ bits to store the original basic block.

The size of the superblock is chosen to be a power of 2 times the size of a basic block.
When calculating the rank, we divide $i$ by first the size of a superblock, and later divide $i$ by the size of a basic block.  This requires two divisions, a potentially-expensive CPU operation.  We reduce this to one division and one shift by setting the size of a superblock to $128$ times the size of a basic block: thus, after we calculate $\lfloor i/63488\rfloor$, we can right shift by $7$ to get $\lfloor i/496\rfloor$.  This choice slightly increases our space, but significantly reduces query time.

 \subsection{SPIDER Select}
\label{sec:spider_select}

Now we describe how SPIDER works on \select{} queries.  We use a rank-based approach, so 
 
select queries also access the rank array and modified bit vector to help guide the query.

\paragraph{Storing Metadata.}
We store two arrays in addition to the rank array and modified bit vector: a high-level select array, and a low-level select array.  
In short, we augment the rank-based select strategy described in Section~\ref{sec:lastmile} with the multi-level strategy.

We define two sampling thresholds, one for each select array: a \defn{high-level sampling threshold} $\sigma_h$, and a \defn{low-level sampling threshold} $\sigma_\ell$.

First, we store a \defn{high-level select array} of length at most $n/63488 + 2$.

We define the high-level sampling threshold to be $\sigma_h = 2^{\lceil\log_2(63488\cdot n_1/n)\rceil}$.  
We will see that the high-level select array has $2 + \lfloor n_1/\sigma_h\rfloor$ entries; thus, $\sigma_h$ is the smallest power of $2$ that ensures that the high-level select array has at most $n/63488 + 2$ entries.\footnote{Choosing $\sigma_h$ to be a power of $2$ allows us to replace a division with a shift, improving query time.}

 The idea of the high-level select array is to store, for every $\sigma_h$th $1$ bit,  the nearest superblock entry of that $1$ bit.  

 The nearest superblock boundary to slot $s'$ is given by $\lfloor s'/63488 + 1/2\rfloor$.
 Thus,  entry $0$ in the high-level select array stores $\lfloor \select{1}/63488 + 1/2\rfloor$.
Then, entry $i$ in the high-level select array (for $i$ from $1$ to $\lfloor (n_1-1)/\sigma_h\rfloor$) stores $\lfloor \select{i\cdot \sigma_h + 1}/63488 + 1/2\rfloor$.  

 Finally, we store $n/63488 - 1$ in the last element of the high-level select array.
 Thus, the high-level select array has $\lfloor (n_1 - 1)/\sigma_h \rfloor + 2$ entries, each
of $64$ bits.  

Then, we store a \defn{low-level select array}.
Our goal is to have each entry in the low-level select array take only $16$ bits---thus, drawing inspiration from the local rank used in Section~\ref{sec:spider_rank}, we define the \defn{superblock offset} of any slot $s'$ to be the number of slots between $s'$ and the first slot of the superblock containing $s'$.  The first slot of the superblock containing $s'$ is $63488\lfloor s'/63488\rfloor$, so the superblock offset of $s'$ is $s' - 63488\lfloor s'/63488\rfloor$.

We define the low-level sampling threshold 
$\sigma_\ell = 2^{\lceil \log_2 (4096\cdot .99n_1/n)\rceil}$.
The low-level select array has $2 + \lfloor n_1/\sigma_\ell\rfloor$ entries, each storing a superblock offset.
Thus, $\sigma_\ell$ is the smallest power of $2$ such that the low-level select array stores at most $2 + n/(.99\cdot 4096)$ entries.\footnote{The $.99$ term is to help performance for bit vectors where very slightly over half of the bits are $1$s: without this term, if (say) $n_1/n = .50001$, we would store roughly $n/8192$ entries (whereas we would store $n/4096$ for $n_1/n = .5$), hurting select query performance.  Thus, the $.99$ term slightly increases our worst-case space bound, but helps performance on datasets with $n_1/n > .5$ and $n_1/n \leq .505$.}

The $i$th entry in the low-level select array stores the superblock offset of $\select{\sigma_\ell \cdot i}$.

In other words, the first slot of the superblock containing $\select{\sigma_\ell \cdot i}$ is slot $s_i = 63488\lfloor\select{\sigma_\ell \cdot i}/63488\rfloor$; thus the $i$th entry stores $\select{\sigma_\ell\cdot i} - s_i$.  
The last entry of the low level array stores $\select{n_1} - 63488\lfloor\select{n_1}/63488\rfloor$.
Each entry has value at most $63488$, so each low-level select array entry can be stored in $16$ bits.

\paragraph{Preprocessing.}

First we build the rank array and modified bit vector in one scan.  Then we can calculate $n_1 = \rank{n-1}$, from which we obtain $\sigma_h$ and $\sigma_\ell$.
Then, in a second scan through the data we build the high level select array and low level select array.

\paragraph{Queries and Predictions.}
On a query $\select{i}$,  we use the high-level select array to find the superblock containing the $i$th set bit as follows.
We first set $s$ to be entry $\lfloor (i-1)/\sigma_h\rfloor$ in the high-level select array; $s$ is a guess for the superblock containing $\select{i}$.

We then do a linear scan starting at $s$ to find the superblock containing $\select{i}$.  We begin with the values stored in the $s$th and $(s+1)$st entries in the rank array; if $i$ is between these values then $\select{i}$ is in the $s$th superblock.  Otherwise, we begin a linear scan for the correct superblock: if $i$ is less than the $s$th rank array entry we decrement $s$ and recurse; if $i$ is greater than the $(s + 1)$st rank array entry we increment $s$ and recurse.  
 At this point, we know that $\select{i}$ is in the $s$th superblock.  

Now we look in the low-level select array.
Let $a$ be the value stored in the $\ell$th  entry of the low-level select array with $\ell = \lfloor (i-1)/\sigma_\ell\rfloor$; let $b$ be the value stored in the $(\ell + 1)$st entry of the low-level select array.  

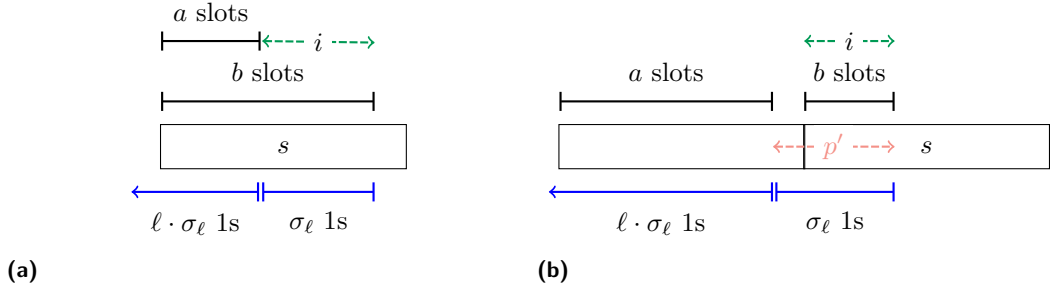
\begin{figure}[ht]
\centering
\begin{subfigure}{.49\textwidth}
\centering
\begin{tikzpicture}
   \node[draw=black,text width=3cm,align=center] at (0,0) {\phantom{|}$s$\phantom{|}};
   \draw[blue,thick, <-|] (-2,-.6) -- node[below=4pt,black]{$\ell\cdot \sigma_\ell$ $1$s} (-.32,-.6);
   \draw[blue,thick, |-|] (-.28,-.6) -- node[below=4pt,black]{$\sigma_\ell$ $1$s} (1.2,-.6);
   \draw[black,thick, |-|] (-1.61,1.4) -- node[above=4pt,black]{$a$ slots} (-.3,1.4);
   \draw[black,thick, |-|] (-1.61,.6) -- node[above=4pt,black]{$b$ slots} (1.2,.6);
   \draw[ForestGreen, <->,thick,dash pattern=on 3pt off 1pt] (-.28,1.4) -- node[midway,black,fill=white,thick]{$i$} (1.2,1.38);
\end{tikzpicture}
\caption{}
\label{fig:well_specified}

\end{subfigure}
\begin{subfigure}{.49\textwidth}
\centering
\begin{tikzpicture}
   \node[draw=black,text width=3cm,align=center] at (0,0) {\phantom{|}\phantom{$s$}\phantom{|}};
   \node[draw=black,text width=3cm,align=center] at (3.22,0) {\phantom{|}$s$\phantom{|}};
   \draw[blue,thick, <-|] (-1.75,-.6) -- node[below=4pt,black]{$\ell\cdot \sigma_\ell$ $1$s} (1.2,-.6);
   \draw[blue,thick, |-|] (1.23,-.6) -- node[below=4pt,black]{$\sigma_\ell$ $1$s} (2.8,-.6);
   \draw[black,thick, |-|] (-1.61,.6) -- node[above=4pt,black]{$a$ slots} (1.2,.6);
   \draw[black,thick, |-|] (1.61,.6) -- node[above=4pt,black]{$b$ slots} (2.8,.6);
   \draw[ForestGreen, <->,thick,dash pattern=on 3pt off 1pt] (1.61,1.4) -- node[midway,black,fill=white,thick]{$i$} (2.8,1.4);
   \draw[Salmon, <->,thick,dash pattern=on 3pt off 1pt] (1.2,0) -- node[midway,fill=white,thick]{$p'$} (2.8,0);
\end{tikzpicture}
\caption{}
\label{fig:not_well_specified}
\end{subfigure}
\label{TODO}
\caption{Two examples of the values used to create a prediction for $\select{i}$.  The example on the left is well-specified, and $i$ must be between $a$ and $b$ in superblock $s$.  The example on the right is not well-specified, since $a$ and $b$ are local ranks from different superblocks.}
\end{figure}

We now create our prediction for $\select{i}$.  
First, some motivation.  
We say that $i$ \defn{well-specified} if 
$a$ and $b$ both represent slots in the $s$th superblock---that is to say, $i$ is well-specified if 
the $s$th superblock contains 
$\select{\sigma_\ell \cdot \ell}$ and 
$\select{\sigma_\ell \cdot (\ell + 1)}$.  See Figure~\ref{fig:well_specified}.
 If $i$ is well-specified, $\select{i}$ must be between $63488\cdot s + a$ and $63488\cdot s + b$.

Let's assume (momentarily) that $i$ is well-specified, and, further, that all $1$ bits are spread evenly between slots 
$63488\cdot s + a$ and $63488\cdot s + b$.  There are $\sigma_\ell$ $1$s between these slots by definition, so if they are spread evenly, there is a $1$ every $(b - a)/\sigma_\ell$ slots.
By definition, there are $i - \sigma_\ell \cdot \ell$ $1$s between $\select{\sigma_\ell \cdot \ell}$ 
and $\select{i}$. 
If these $1s$ are evenly spread, 
substituting $\select{\sigma_\ell\cdot \ell} =  63488 \cdot s + a$,
then $\select{i}$ is located at position
\vspace{-.05in}
\begin{equation}
\label{eq:prediction}
p = 63488\cdot s + a + \frac{b - a}{\sigma_\ell}\cdot(i - \sigma_\ell\cdot \ell).
\end{equation}

\vspace{-.05in}
Thus, for any $i$, we define $p$ using Equation~\ref{eq:prediction}; we search for $\select{i}$ using a linear scan beginning at $p$.  Specifically, since $\select{i}$ is contained in the $s$th superblock, we begin at basic block $B = \lfloor p/496\rfloor$.  We proceed with a linear scan.
First, we ensure that the number of $1$ bits before basic block $B$ is at most $i$.

The number of $1$ bits before block $B$ can be obtained by summing the local rank of $B$ with the $\lfloor B/128\rfloor$th entry in the rank array.  While this sum is at most $i$, we decrement $B$ and check again.  Then, we use the same strategy to ensure that the number of $1$ bits before block $B+1$ is at least $i$ (while it is not, we increment $B$).  After both loops complete, we know that $\select{i}$ is in block $B$.

Now, let us revisit the assumptions we made when defining predictions: what happens when $i$ is not well defined?  
Even on very well-behaved data, our low-level select array entries will cross superblock boundaries relatively frequently (see Figure~\ref{fig:not_well_specified}); we must ensure that in this case our queries are correct, and that our predictions are as accurate as possible.

At first glance, correctness appears to be guaranteed since we do a linear scan for the correct basic block---however, our algorithm as presented is not correct if $b < a$.  In particular, 
if $b < a$ and $i$ is in the first or last superblock, we may have $p < 0$ or $p > n-1$.  
 On top of the correctness issue, $b < a$ also causes particularly bad predictions.  We found that this case is common enough to noticeably slow down our queries.

Thus, if $b< a$, we make the following adjustment to ensure correctness and improve the solution quality. In short, we can test which sampled $1$ is in the same superblock as $i$, and adjust $a$ or $b$ accordingly.  Specifically, we test if entry $s$ in the rank array is at least $\sigma_\ell\cdot \ell$.  If so, $\select{\sigma_\ell\cdot \ell}$ is in a superblock before $i$, so $a$ is not in the same superblock as $i$; we calculate $a' = a - 63488$.  Otherwise we set $b' = b + 63488$.  In each case we calculate $p'$ using Equation~\ref{eq:prediction} with $a'$ or $b'$, and begin the search at $B= \lfloor p'/496\rfloor$.  For example, in Figure~\ref{fig:not_well_specified}, $a'$ would be a negative number representing the slots between $a$ and the beginning of superblock $s$; thus, $p'$ is in the range shown in the figure.

This fix improves results in a good prediction if the sampled bits are in successive blocks.  
Furthermore, a case-by-case analysis shows that $p$ is now always between $0$ and $n-1$.  Of course, this simple fix does not always result in a good prediction---if $a$ and $b$ are separated by multiple blocks, or if $i$ is not well-specified but $a < b$, then $p$ is unlikely to be accurate---but these cases are rare enough on real-world data as to not significantly impact performance.

After the above, we have found the basic block $B$ containing $\select{i}$.
We perform a select within $B$ by first masking out the first 16 bits, and then using fast select (Section~\ref{sec:lastmile}).

\paragraph{Space.}
The rank array requires $64\cdot n/63488$ bits.  The modified bit vector requires $512\cdot n/496$ bits.  Summing, we obtain $3.33\%$ extra space.
 The high-level select array requires at most $64\cdot n/63488$ bits; the low-level select array requires at most $16\cdot n/4096/.99$ bits.  Thus, select requires $.495\%$ space.
 Summing, SPIDER requires a  $3.83\%$ space overhead.

\subsection{Non-Interleaved SPIDER}
\label{sec:ni-spider}

For some use cases, it may not be possible to modify $V$.  With these cases in mind, we define a version of spider that does not interleave metadata: it leaves the original bit vector untouched, and uses 3.62\% space to help answer rank and select queries.  We call this data structure Non-Interleaved SPIDER, or \texttt{ni-spider}.

Non-Interleaved SPIDER has further advantages, even for use cases where the bit vector may be modified.

It is simpler: \texttt{ni-spider} uses a simpler one-level select method that is easier to implement, and it avoids several corner cases.  Furthermore, we discuss in Section~\ref{sec:comparison} that the non-interleaved rank metadata 

improves select performance on some datasets.

\paragraph{Non-Interleaved Rank Data Structure.}
 The rank data structure and queries for non-interleaved SPIDER is essentially the same as \texttt{pasta-wide} as given by by Kurpicz~\cite{Kurpicz22} (however, the two data structures differ for select queries).  Nonetheless, we describe the rank data structure for completeness, as well as for reference when describing select queries.

 We partition the array into $n/65536$ contiguous superblocks of $65536$ bits.  The \defn{rank array} 
 consists of $n/65536$ $64$-bit entries, where the $i$th entry stores $\rank{65536i-1}$.  
 
 Then, we store a \defn{second level rank array}.  
 We partition the bit vector into $n/512$ basic blocks of $512$ bits.

 We define the \defn{local rank} of block $i$ to be 

 $\rank{512i - 1} - \rank{65536\lfloor 512i/65536\rfloor - 1}$.  We store this value in slot $i$ of the second level rank array.  We retain $V$ as well to help with rank and select queries.

 \paragraph{Rank Queries.}
To find \rank{i}, we begin by using the rank array to find $r_1 = \rank{\lfloor i/65536\rfloor - 1}$ (this can be done by looking up the $ \lfloor i/65536 \rfloor$th position in the rank array). 
 Then, we use the second level rank array to find $r_2 = \rank{\lfloor i/512\rfloor - 1}$.

 Finally, we use \texttt{popcount} to  count the number of $1$s before position $i$ within basic block $b_2$ of $V$ (see Section~\ref{sec:lastmile}).

\paragraph{Non-interleaved Select Data Structure}
For non-interleaved SPIDER we use a single-level select array.  We discuss why a single level is sufficient in Section~\ref{sec:comparison}.  

We calculate a single \defn{sampling threshold} $\sigma=2^{\lceil\log_2 (16384\cdot n_1/n)\rceil}$.  The \defn{select array} has at most $n/16384$ entries, each of $64$ bits.\footnote{The constant $16384$ was chosen to remain under 4\% space usage.}  Entry $i$ of the select array stores $\select{\sigma\cdot i}$.

\paragraph{Non-interleaved Select Queries}
To answer $\select{i}$, let $a$ be the value stored in the $\lfloor i/\sigma\rfloor$th entry of the select array, and let $b$ be the value stored in the $(\lfloor i/\sigma\rfloor + 1)$th entry.  Thus we predict the value of $\select{i}$ to be
    $p = a + \frac{b - a}{\sigma}(i - \sigma\lfloor i/\sigma\rfloor)$.

Since we store a single select array, there is no issue when $a$ and $b$ span a superblock boundary.

We begin at basic block $B = \lfloor p/512\rfloor$.  
We use the rank array and the second-level rank array to scan for the correct basic block.
The number of $1$s before $B$ can be obtained by adding the $\lfloor B/128\rfloor$th entry in the rank array to the $B$th entry in the second-level rank array; while this number is greater than $i$ we decrement $B$.  We then increment $B$ while the $(\lfloor B/128\rfloor - 1)$st rank array entry plus the $(B+1)$st second-level rank array entry are smaller than $i$.
 Once $B$ is found, we select in a block using fast select (see Section~\ref{sec:lastmile}).  

\paragraph{Space Usage.} Summing ${64n}/{65536} + {16n}/{512} + {64n}/{16384}$ we obtain $3.62\%$ extra space.

\subsection{Comparing SPIDER and Non-Interleaved SPIDER Select Queries}
\label{sec:comparison}

Here we discuss some performance differences between SPIDER and Non-Interleaved SPIDER.  

For rank queries, the interleaved strategy has strictly better cache performance.  SPIDER incurs $\leq 1$ cache miss when accessing the rank array, and a cache miss when accessing the modified bit vector.  Non-Interleaved SPIDER incurs $\leq 1$ cache miss for the rank array, a cache miss to access the second-level rank array, and another to access the bit vector.

Interleaving the ranks has an immediate impact on select queries.  Non-Interleaved SPIDER does a linear search for the block containing $\select{i}$ in the second-level rank array, whereas SPIDER searches the modified bit vector.  

 This leads to an interesting tradeoff in cache performance.  SPIDER incurs a cache miss each time it considers a new basic block.  In contrast, Non-Interleaved SPIDER almost never incurs more than one cache miss in the second-level rank array, since $32$ consecutive basic blocks fit in a cache line.  However, Non-Interleaved SPIDER incurs an extra cache miss to access the bit vector itself.  Thus, assuming that all array accesses result in a cache miss, if $p$ is $g$ basic blocks away from $\select{i}$, SPIDER incurs $1 + g$ cache misses.  Non-Interleaved SPIDER incurs on average $2 + g/16$ cache misses.  

 Thus, the cache advantages of SPIDER's rank queries also apply to select queries if the predictions give exactly the correct basic block.  However, inaccurate predictions immediately lead to cache misses for SPIDER, whereas Non-Interleaved SPIDER's cache performance is minimally affected by prediction quality.  The break-even point is when each prediction is $1$ basic block away from the correct basic block on average.

This tradeoff motivates our two-level select array for SPIDER.  The two-level design leads to more computation and a more complicated data structure.  In exchange, the two-level array samples far more $1$ positions in the same space, increasing the prediction quality.  This is particularly important on sparse data (see Figure~\ref{fig:spider_comparison}).

In addition to the difference in cost for bad predictions, SPIDER has a secondary advantage from avoid a second-level rank array: the extra array uses up the cache, potentially causing other metadata to be evicted much earlier.  This may cause Non-Interleaved SPIDER to incur more cache misses than SPIDER when $n$ is small.

\subsection{SPIDER vs Non-Interleaved SPIDER Predictions}

  We ran experiments to compare SPIDER and Non-Interleaved SPIDER in more detail. 
 In Table~\ref{table:guess_accuracy}, we give results for the accuracy of the prediction of each method: the table gives the average number of incorrect basic blocks looked at by SPIDER and Non-Interleaved SPIDER on three datasets.  
 We see that the predictions are generally very accurate; even on the sparse Protein dataset, the predictions are off by approximately one basic block.

 We also see that, as one might expect, real data is more difficult to predict than random synthetic data.  Nonetheless, the results in Figure~\ref{fig:select} show that even with this modest decrease in prediction quality, the algorithms retain good results.

\section{Experiments}%
\label{sec:experiments}
 We ran all experiments on a x86-64 11th Gen Intel(R) Core(TM) i7-1165G7 @ 2.80GHz with L1d, L1i, L2, and L3 cache sizes 192 KiB, 128 KiB, 5 MiB and 12 MiB respectively. The machine had 32GB RAM and was running Ubuntu 20.04 LTS. The code was written in \texttt{C} and compiled using gcc version 9.4.0 with compiler flags: \texttt{-O9 -march=native -mpopcnt -mlzcnt}. 

For all experiments, we first ran $10^8$ warmup queries to ensure a warm cache.  Then, we ran $10^8$ queries, recording the total time 
and dividing to get the average query time.
  We ran each experiment $5$ times; the results are the average of these $5$ experiments.

We give the code for all \texttt{spider} variants at \url{https://github.com/williams-cs/spider}.

\paragraph{Datasets.}
Our experiments use both synthetic and real data sets.
For synthetic data, we provide three different densities: 10\%, 50\%, and 90\%. The data is generated by setting each bit to 1 independently with probabilities 0.1, 0.5, or 0.9 respectively. We refer to these datasets as 10\% Random, 50\% Random, and 90\% Random.

We give two bit vectors based on protein data, which we call Protein and Protein-Even.
These bit vectors are based on the Uniref90 protein data set~\cite{SuzekHuMc07}.\footnote{Obtained from \url{https://www.uniprot.org/help/uniref}.}   To generate Protein, we set each occurrence of a leucine amino acid (character \texttt{L}) to a 1 bit and all others to zero. 
Protein is sparse: 9.72\% of the bits are $1$s, and the rest are $0$s.  
We also give experiments on a second dataset, which we call Protein-Even; this dataset was generated by setting characters \texttt{A} through \texttt{L} to $0$ and all others to $1$; this gives a density of $44.5\%$ $1$s.

The second dataset consists of a dump of all Wikipedia articles.\footnote{We used the Wikipedia dump from 01-01-2024: \texttt{enwiki-20240101-pages-articles-multistream.xml}.  This was downloaded from \url{https://mirror.accum.se/mirror/wikimedia.org/dumps/enwiki/}, and extracted from xml to text using WikiExtractor~\cite{Attardi15}.}
We mapped all characters \texttt{a} through \texttt{n} and \texttt{A} through \texttt{N} to 1 and all others to $0$; this resulted in a dataset with 47\% 1s.
The Wikipedia data has less than $32$ billion characters, so results only go up to $16$ billion.

We note that wavelet trees in particular motivate our Wikipedia and Protein-Even datasets: the top level in a wavelet tree is a bit vector that maps (roughly) half the original characters to $1$ and the other half to $0$~\cite{Navarro14}.  Thus, our Wikipedia and Protein-Even bit vectors are exactly the top level of a wavelet tree on the original wikipedia and Uniref 90 datasets.

\begin{figure}[t]
  \centering
  \begin{subfigure}{.5\textwidth}
    \centering
    \includegraphics[width=1\linewidth]{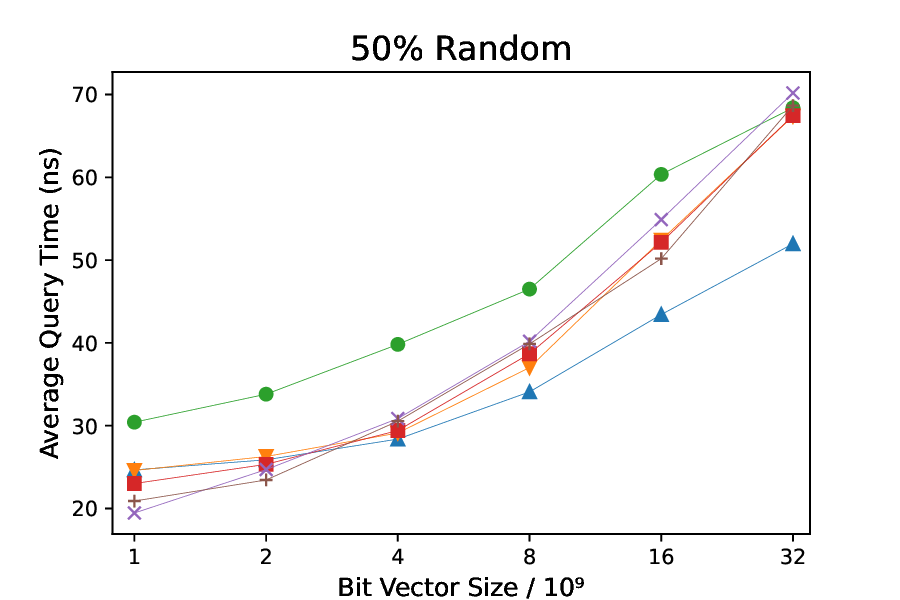}
    \label{fig:rank-sub4}
  \end{subfigure}%
  \begin{subfigure}{.5\textwidth}
    \centering
    \includegraphics[width=1\linewidth]{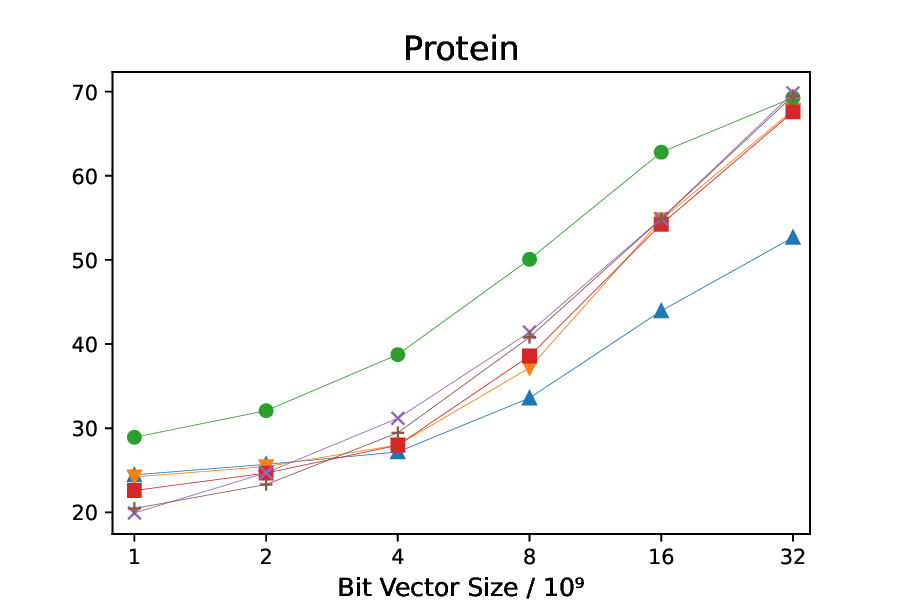}
    \label{fig:rank-sub2}
  \end{subfigure}
  \begin{subfigure}{\textwidth}
  \vspace{-.08in}
    \centering
    \includegraphics[width=0.5\linewidth]{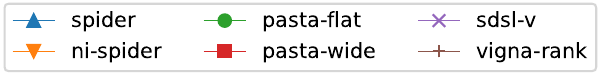}
    \label{fig:rank-legend}
  \vspace{-.08in}
  \end{subfigure}
  \caption{Average rank query time in nanoseconds, for bit vectors from 1 billion to 32 billion bits.}
  \label{fig:rank}
\end{figure}

 \begin{figure}
  \centering
  \begin{subfigure}{.49\textwidth}
    \centering
    \includegraphics[width=1\linewidth]{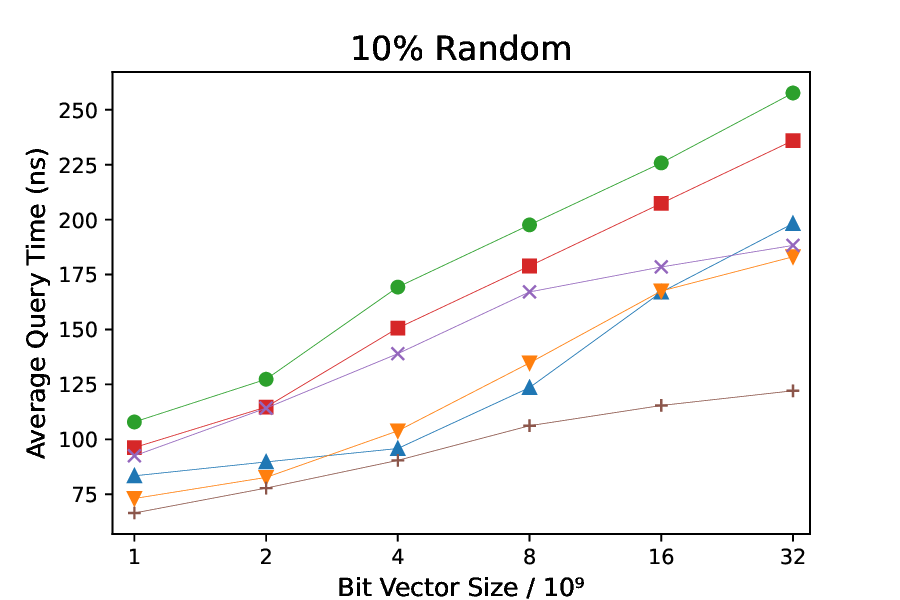}
    \label{fig:select-sub1}
  \end{subfigure}%
  \begin{subfigure}{.49\textwidth}
    \centering
    \includegraphics[width=1\linewidth]{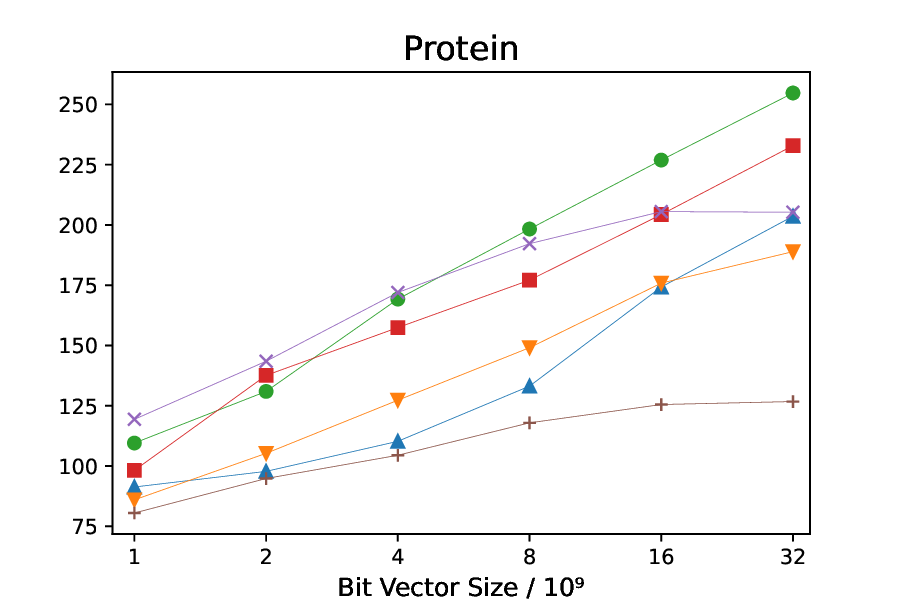}
    \label{fig:select-sub2}
  \end{subfigure}\\
  \vspace{-.2in}
  \begin{subfigure}{.49\textwidth}
    \centering
    \includegraphics[width=1\linewidth]{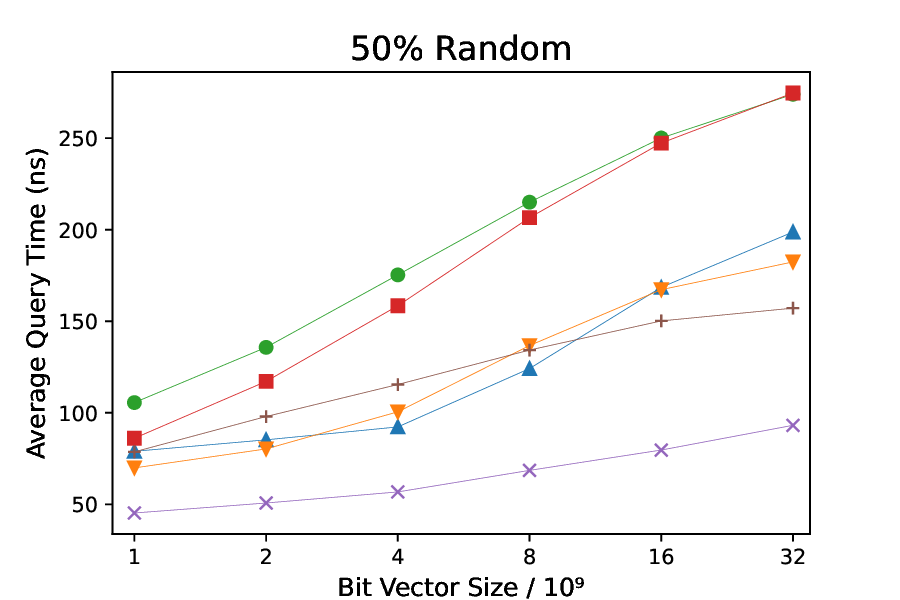}
    \label{fig:select-sub3}
  \end{subfigure}%
  \begin{subfigure}{.49\textwidth}
    \centering
    \includegraphics[width=1\linewidth]{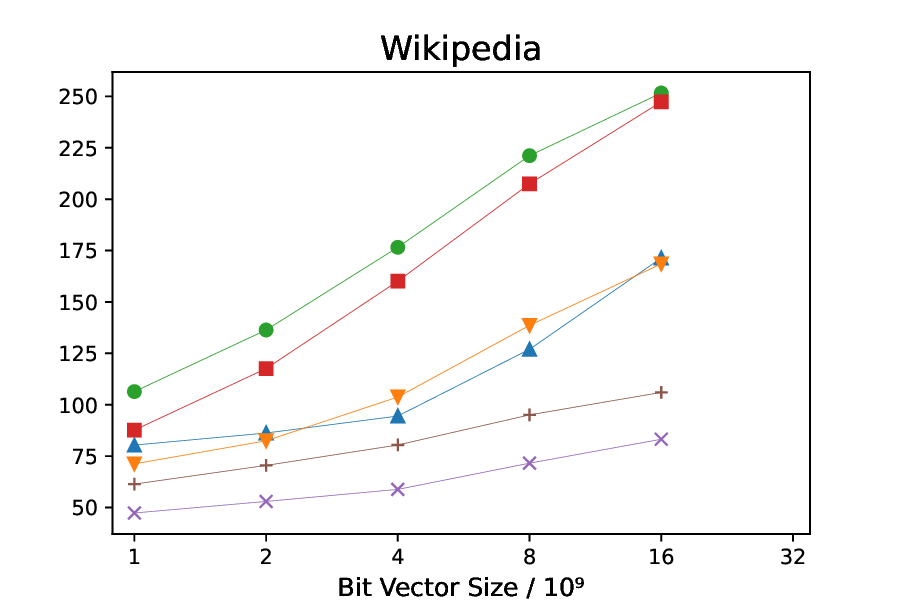}
    \label{fig:select-sub4}
  \end{subfigure}\\
  \vspace{-.2in}
  \begin{subfigure}{.49\textwidth}
    \centering
  \includegraphics[width=\textwidth]{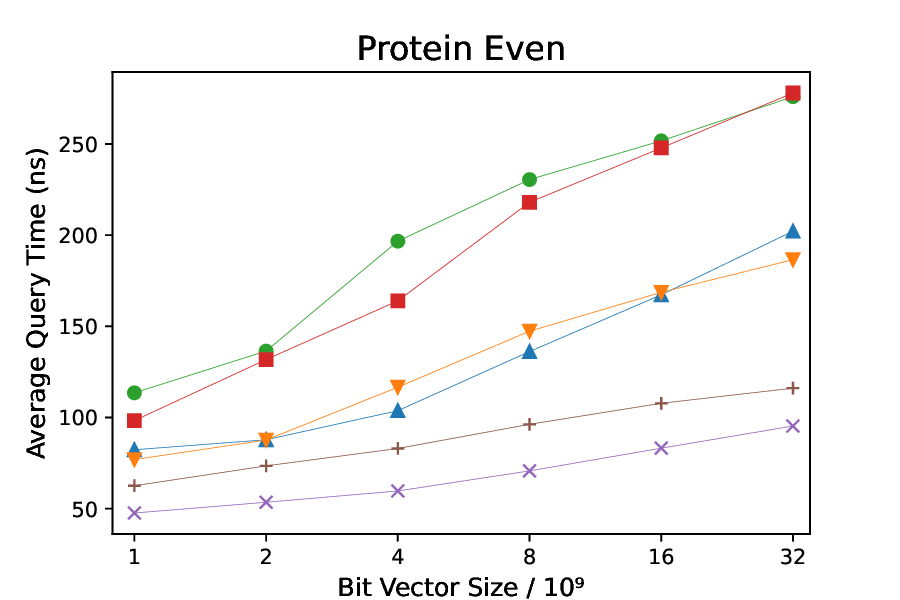}
  \end{subfigure}%
    \begin{subfigure}{.49\textwidth}
    \centering
  \includegraphics[width=\textwidth]{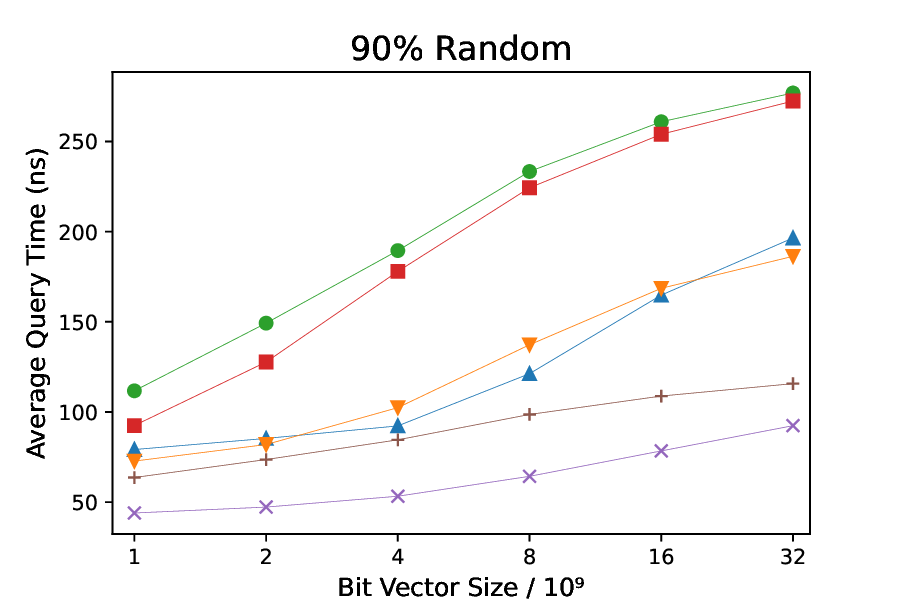}
  \end{subfigure}\\
  \vspace{.05in}
  \begin{subfigure}{.99\textwidth}
    \centering \includegraphics[width=0.6\linewidth]{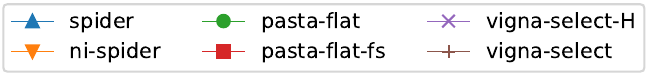}
    \label{fig:select-legend}
  \end{subfigure}
  \caption{Average select query time in nanoseconds, for datasets from 1 billion to 32 billion bits.}
  \label{fig:select}
\end{figure}
\paragraph{Rank Query Performance.}
We compare \texttt{spider} and \texttt{ni-spider} to four other datasets.  First, we compare to \texttt{pasta-flat}: this is the most performant data structure that can handle both rank and select queries with under 5\% space.  We also compare to \texttt{pasta-wide}, \texttt{sdsl-v}, and \texttt{vigna-rank}, all of which are tuned specifically for rank queries and cannot handle select queries efficiently (see Section~\ref{sec:related_state_of_the_art}).
We note that \texttt{ni-spider} and \texttt{pasta-wide} are effectively the same method for rank (though not for select) queries, as are \texttt{sdsl-v} and \texttt{vigna-rank}.  Unsurprisingly, their performance values are very similar; any discrepancies are likely due to small implementation differences.
Nonetheless, we include all results for completeness.

We give our rank results in Figure~\ref{fig:rank}; results for the remaining datasets are essentially identical and are omitted.
 Overall, SPIDER gives the best rank results for large $n$, even compared to data structures that have worse space efficiency and are tailored specifically to rank queries.  
For smaller $n$, SPIDER is still competitive, though is not the best, likely due to the extra computation required to extract the interleaved metadata.

\paragraph{Select Query Performance.}
We compare \texttt{spider} and \texttt{ni-spider} to 
\texttt{pasta-flat} and
\texttt{pasta-flat-fs}, the most performant data structures under 10\% space,
 and \texttt{vigna-select-H} and \texttt{vigna-select},  which use over 10\% space and cannot answer rank queries.

We give our select results in Figure~\ref{fig:select}.
SPIDER achieves better performance than \texttt{pasta-flat}, the state of the art data structure using less than 5\% space.
Even compared to the space-inefficient data structures, SPIDER performs quite well, achieving roughly similar performance to \texttt{vigna-select} on 50\% random and 90\% random data, and performing only slightly worse than \texttt{vigna-select} on all other datasets.  SPIDER does noticeably worse than \texttt{vigna-select-H} on Wikipedia, Protein Even, and Random 50\% data, but better on the sparse datasets---this is unsurprising since \texttt{vigna-select-H} is specifically tailored to perform well for data sets with $n_1/n\approx .5$.

SPIDER and Non-Interleaved SPIDER achieve similar select performance on all datasets.  

\begin{figure}[t]
 \begin{minipage}{.45\textwidth}
  \centering
  \includegraphics[width=\textwidth]{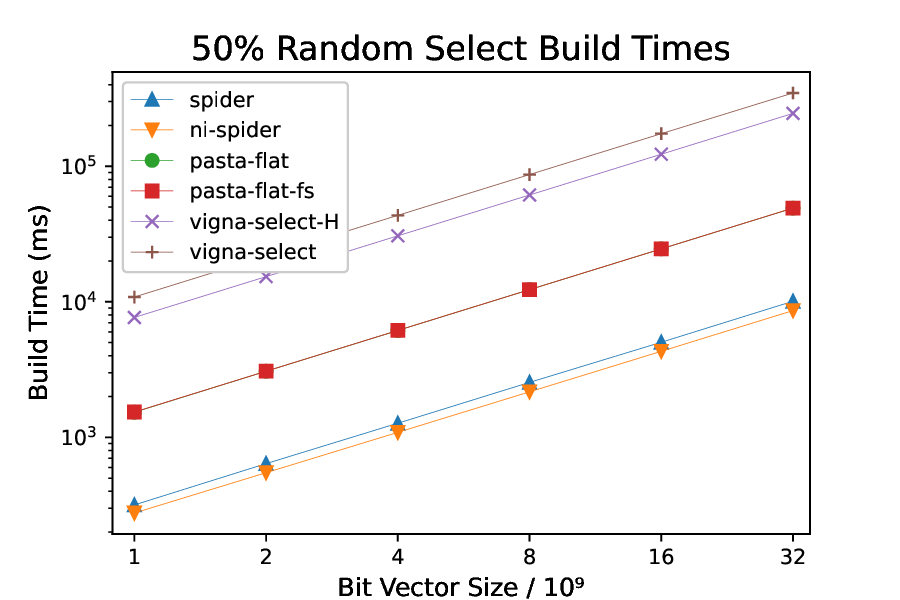}
  \captionof{figure}{Total build time in milliseconds.  The $y$ axis uses a logarithmic scale.}
  \label{fig:buildtime}
  \end{minipage}
  \hfill
  \begin{minipage}{.45\textwidth}
    \centering
     \begin{tabular}{c|c c c c} 
         Dataset & \texttt{spider} & \texttt{ni-spider} \\ 
         \hline
        50\% Random & 0.040526 & 0.078617 \\
        Protein & 0.413667 & 1.387546 \\
        Wikipedia & 0.059781 & 0.156082\\
     \end{tabular}
     \vspace{.1in}
    \captionof{table}{Prediction accuracy comparison of SPIDER and Non-Interleaved SPIDER.  All datasets have 1 billion bits.  The values given are the average number of incorrect basic blocks searched during a select query before the correct basic block is found.}
    \label{table:guess_accuracy}
    \end{minipage}
\end{figure}

\paragraph{Build Time.}
We ran experiments comparing build time.  Preprocessing is very lightweight for SPIDER and Non-Interleaved SPIDER: the data structures only consist of a few arrays that can each be built with a linear scan through the data.  
We used \texttt{popcount} and fast select during our build to minimize the computation time.

We give the results for build on 50\% Random data in Figure~\ref{fig:buildtime}; SPIDER and Non-Interleaved SPIDER are significantly faster to build than any other data structure.  We note that the \texttt{pasta-flat} and \texttt{pasta-flat-fs} lines overlap in this figure.

\begin{figure}[t]
  \centering
  \begin{subfigure}{.5\textwidth}
    \centering
    \includegraphics[width=1\linewidth]{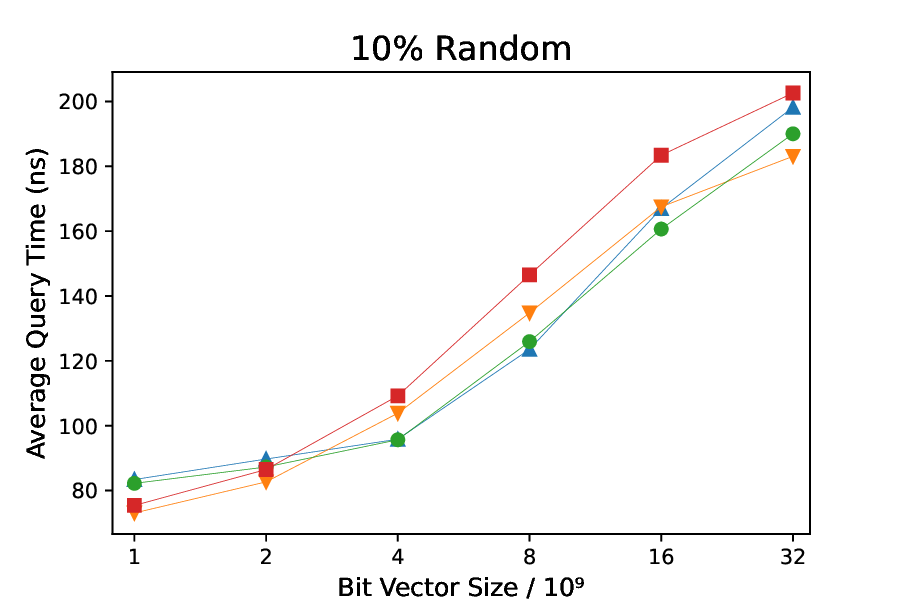}
    \label{fig:sub1}
  \end{subfigure}%
  \begin{subfigure}{.5\textwidth}
    \centering
    \includegraphics[width=1\linewidth]{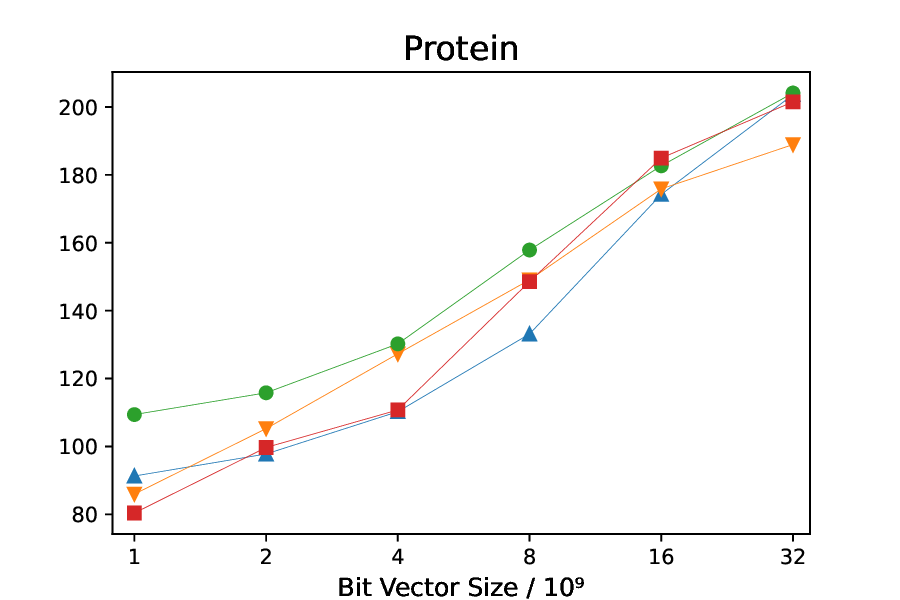}
    \label{fig:sub2}
  \end{subfigure}
  \begin{subfigure}{.5\textwidth}
    \centering
    \includegraphics[width=1\linewidth]{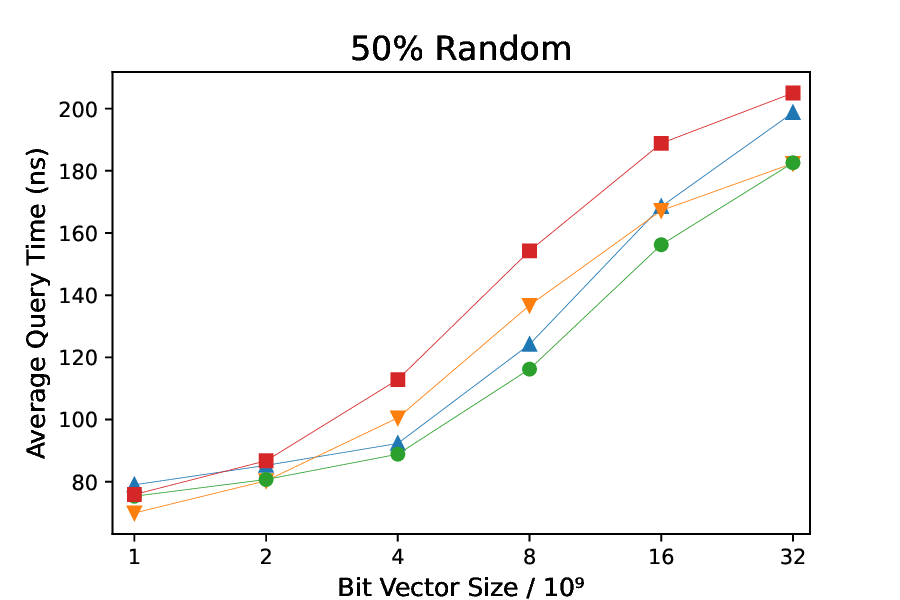}
    \label{fig:sub3}
  \end{subfigure}%
  \begin{subfigure}{.5\textwidth}
    \centering
    \includegraphics[width=1\linewidth]{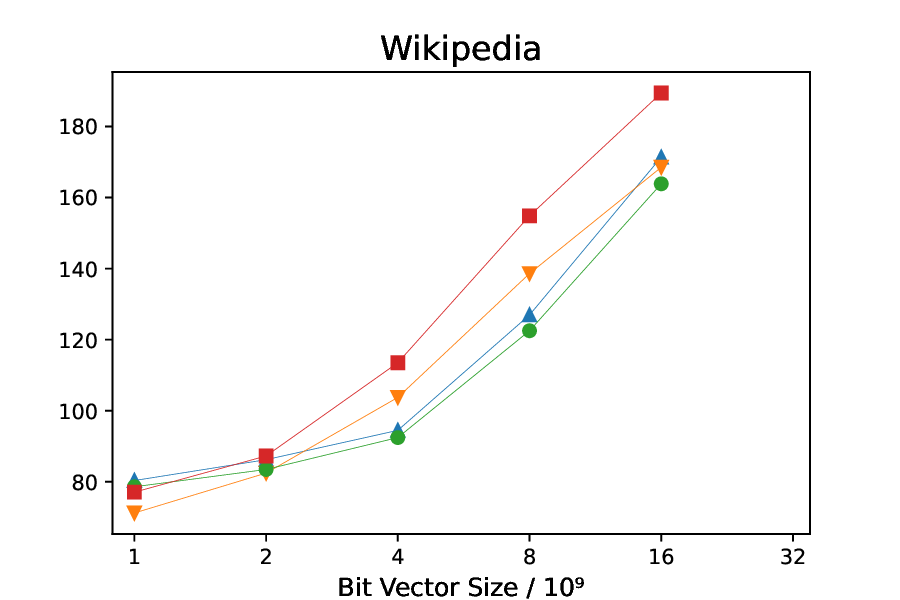}
    \label{fig:sub4}
  \end{subfigure}
  \begin{subfigure}{\textwidth}
    \centering
    \includegraphics[width=0.9\linewidth]{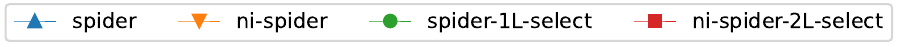}
    \label{fig:sub5}
  \end{subfigure}
  \caption{Average time for a select query in nanoseconds for the four different variants of SPIDER.}
  \label{fig:spider_comparison}
\end{figure}

 \paragraph{Comparison of SPIDER Variants.}
 SPIDER and Non-Interleaved SPIDER differ in two ways: Non-Interleaved SPIDER does not interleave data, and uses a single-level select array.  Here, we compare to the other two natural variants of SPIDER: \texttt{spider-1L-select} interleaves the ranks and keeps a $1$-level select array, and \texttt{ni-spider-2L-select} does not interleave ranks, but has a $2$-level select array.  The results are given in Figure~\ref{fig:spider_comparison}.

Overall, SPIDER and Non-Interleaved SPIDER have the best results.  However, the other two variants often achieve similar or even non-negligibly better performance, particularly for small values of $n$.  One interesting point is that \texttt{spider-1L-select} has a single-level select array, but interleaves local ranks (and thus has a high cost for bad guesses)---for the Random 50\% data the single-level select array is sufficient for high-quality predictions, whereas for the sparse, relatively-difficult-to-predict Protein bit vector it lags noticeably behind the other variants.

We note that the distinction between the $1$-level and $2$-level select arrays has no impact on rank performance.  Therefore, the performance of \texttt{spider-1L-select} is identical to that of \texttt{spider} (and \texttt{ni-spider-2L-select} is identical to \texttt{ni-spider}) in Figure~\ref{fig:rank}, so we do not give additional rank experiments here.

\section{Conclusion}
\label{sec:conclusion}

In this work we give SPIDER, a succinct data structure for rank and select.  SPIDER interleaves metadata with the underlying bit vector to improve rank performance, and uses predictions and a two-level select array to improve select performance, while using only 3.83\% additional space.  For rank queries, SPIDER gives an up to 
22\%

improvement in query speed over the state of the art, even comparing to data structures using much more space.  For select queries, SPIDER is up to
41\%

faster than the state of the art when comparing only to data structures that use under 5\% space. SPIDER is as little as 
3.1\%

slower than the state of the art data structure (which uses 12.2\% extra space and cannot answer rank queries).  
While these speedups represent our best results, SPIDER consistently outperforms data structures using similar space, and is competitive with state of the art data structures using more space.
Thus, SPIDER makes significant progress on eliminating the tradeoff between space usage and query time for rank and select queries.  
We show that SPIDER is effective on both real and synthetic datasets.

Our second data structure, Non-Interleaved SPIDER, leaves the underlying bit vector unchanged.  Non-Interleaved SPIDER roughly matches the performance of the best known rank data structure other than SPIDER, and matches SPIDER select performance.

\bibliography{spider}

\end{document}